\documentclass[12pt]{article}
\usepackage{a4wide,epsfig,psfrag,amsmath,amssymb,cite,scalefnt}
\usepackage{color}

\usepackage{amsmath}
\usepackage{graphicx}

\usepackage{color}

\renewcommand{\d}{\mathrm{d}}

\renewcommand{\H}[2]{\textrm{H}_{#1}(#2)} 

\newcommand{\ep}{\epsilon}

\newcommand{\dd}{\mathrm{d}}

\newcommand{\gsim}{\;\rlap{\lower 3.5 pt \hbox{$\mathchar \sim$}} \raise 1pt
 \hbox {$>$}\;}
\newcommand{\lsim}{\;\rlap{\lower 3.5 pt \hbox{$\mathchar \sim$}} \raise 1pt
 \hbox {$<$}\;}

\allowdisplaybreaks

\begin{document}


\title{\vskip-3cm{\baselineskip14pt
    \begin{flushleft}
      \normalsize SFB/CPP-12-39\\
      \normalsize TTP12-20
  \end{flushleft}}
  \vskip1.5cm
  Moments of heavy quark correlators with two
  masses: exact mass dependence to three loops
}

\author{
  Jonathan~Grigo, Jens~Hoff, Peter~Marquard, Matthias~Steinhauser
  \\
  {\small\it Institut f{\"u}r Theoretische Teilchenphysik, }
  {\small\it Karlsruhe Institute of Technology (KIT)}\\
  {\small\it 76128 Karlsruhe, Germany}
}

\date{}

\maketitle

\thispagestyle{empty}

\begin{abstract}
  We compute moments of non-diagonal correlators with two massive quarks.
  Results are obtained where no restriction on the ratio of the masses is
  assumed. Both analytical and numerical methods are applied in order to
  evaluate the two-scale master integrals at three loops. We provide explicit
  results for the latter which are useful for other calculations.  As a
  by-product we obtain results for the electroweak $\rho$ parameter up to
  three loops which can be applied to a fourth generation of quarks with
  arbitrary masses.

  \medskip


\end{abstract}


\newpage


\section{QCD corrections to current correlators}
\label{sec:introduction}

In recent years many multi-loop results to current correlators became
available. In the massless approximation the state-of-the art are four-loop
integrals which have been applied to $\tau$ and $Z$-boson decays in order to
extract $\alpha_s$~\cite{Baikov:2008jh,Baikov:2012er}. On the other hand, in
the limit of small external momentum the so-called moments have been obtained
also up to four-loop
order~\cite{Kallen:1955fb,Chetyrkin:1996cf,Chetyrkin:1997mb,Maier:2007yn,Chetyrkin:2006xg,Boughezal:2006px,Sturm:2008eb,Maier:2008he,Kiyo:2009gb}
and both $\alpha_s$ and the charm and bottom quark masses could be extracted
by comparing to experimental data or lattice
simulations~\cite{Kuhn:2007vp,Allison:2008xk,Chetyrkin:2009fv,McNeile:2010ji,Chetyrkin:2010ic}.
The currents involved in these calculations only couple to one quark flavour.

As far as the non-diagonal correlators, where the external currents couple to
two quark flavours with different masses, are concerned results for the
moments up to three loops are known for the limit where one mass is
zero~\cite{Djouadi:1993ss,Djouadi:1994gf,Chetyrkin:2000mq,Chetyrkin:2001je,Maier:2011jd,Hoff:2011ge}.
At four-loop order the vector and axial-vector correlators have only been
computed for vanishing external momentum (see
Ref.~\cite{Chetyrkin:2006bj,Boughezal:2006xk}) in order to obtain four-loop
corrections to the electroweak $\rho$ parameter in the Standard
Model~\cite{Schroder:2005db,Chetyrkin:2006bj,Boughezal:2006xk}.

It has been argued in Ref.~\cite{:2010jy} that it would be desirable to have
at hand moments of the heavy-light currents also for general values of the two
quark masses.  The reason to not only consider physical mass values of the
involved quarks is related to lattice simulations where often a variety of
different masses are considered in order to be able to extrapolate to the mass
values of physical interest.  In Ref.~\cite{Hoff:2011ge} this task has been solved
with the help of expansions around the equal-mass case and around the limit
where one mass is much smaller than the other. In this paper we present exact
results up to three loops and thus improve the findings of~\cite{Hoff:2011ge}.

We adopt the notation from Ref.~\cite{Hoff:2011ge}. Let us nevertheless
present the relevant formulas in order to have a self-contained
presentation of the results.

The momentum-space correlator formed by a vector ($v$), axial-vector ($a$),
scalar ($s$) or pseudo-scalar ($p$) current is given by
\begin{eqnarray}
  \left(-q^2g_{\mu\nu}+q_\mu q_\nu\right)\,\Pi^\delta(q^2)
  +q_\mu q_\nu\,\Pi^\delta_L(q^2)
  &=&
  i\int {\rm d}x\,e^{iqx}\langle 0|Tj^\delta_\mu(x) j^{\delta\dagger}_\nu(0)|0 \rangle\,,
  \nonumber 
  \\
  q^2\,\Pi^\delta(q^2)
  &=&
  i\int {\rm d}x\,e^{iqx} \langle 0|Tj^\delta(x)j^{\delta\dagger}(0)|0 \rangle
  \,,
  \label{eq::pispdef}
\end{eqnarray}
with
\begin{eqnarray}
  j_\mu^v = \bar{\psi}_1\gamma_\mu\psi_2\,,\quad
  j_\mu^a = \bar{\psi}_1\gamma_\mu\gamma^5\psi_2\,,\quad
  j^s     = \bar{\psi}_1\psi_2\,,\quad
  j^p     = \bar{\psi}_1 i\gamma^5 \psi_2
  \,.
  \label{eq::currents}
\end{eqnarray}
In our calculations we are allowed to use anti-commuting $\gamma_5$
since for $\psi_1\not=\psi_2$ only non-singlet diagrams contribute.

We work in $n_f$-flavour QCD with two massive and $n_f-2$ massless quarks.
In our final results we adopt the $\overline{\rm MS}$ renormalization scheme
both for the parameters (strong coupling constant $\alpha_s$
and the quark masses $m_1$ and $m_2$) and the overall renormalization
of the correlator which in the following is indicated by a bar.

In the limit $q^2\to0$ the quantity $\bar\Pi^\delta(q^2)$ (and analogously
$\bar\Pi_L^\delta(q^2)$) can be cast in the form
\begin{eqnarray}
  \bar{\Pi}^\delta(q^2) &=& \frac{3}{16\pi^2}
  \sum_{n\ge-1} \bar{C}^\delta_n(x) z^n
  \,,
\end{eqnarray}
where the dimensionless variables\footnote{In the following we do not
  write $m_1^{(n_f)}(\mu)$ and $m_2^{(n_f)}(\mu)$ but suppress the dependence
  on the renormalization scale $\mu$ and the number of flavours $n_f$.}
\begin{eqnarray}
  z = \frac{q^2}{m_1^2}\,,\qquad\mbox{and}\qquad  x = \frac{m_2}{m_1}\,,
\end{eqnarray}
have been introduced. Note that we assume $0\le x\le1$. Results for $m_2>m_1$
are easily obtained by interchanging $m_1$ and $m_2$.  We refer to the
coefficients $\bar{C}^\delta_n(x)$ also as moments. Their perturbative
expansion can be written as
\begin{eqnarray}
  \bar{C}^\delta_n = \bar{C}^{(0),\delta}_n
  + \frac{\alpha_s}{\pi} \bar{C}^{(1),\delta}_n
  + \left(\frac{\alpha_s}{\pi}\right)^2 \bar{C}^{(2),\delta}_n
  + \ldots
  \,,
\end{eqnarray}
where the arguments $x$ and $\mu$ (renormalization scale) are suppressed.
Note that we have the symmetry relations $\bar{C}^s_n(x)=\bar{C}^p_n(-x)$ and
$\bar{C}^v_n(x)=\bar{C}^a_n(-x)$ (see, e.g., Ref.~\cite{Chetyrkin:1996ia})
which we use to cross check our results. A further check is provided by
computing the moments of the longitudinal contribution $\Pi_L^{v,a}(q^2)$ and
compare the result of the moments of $\Pi^{s,p}(q^2)$ with the help of the
relations
\begin{eqnarray}
  \bar{C}_{L,n}^v &=& (1-x)^2 \bar{C}_{n+1}^s\,,\nonumber\\
  \bar{C}_{L,n}^a &=& (1+x)^2 \bar{C}_{n+1}^p\,.
\end{eqnarray}

For completeness, let us mention that the
QED-like normalization of the polarization function with
$\Pi^\delta(0)=0$ can be obtained from $\bar{\Pi}^\delta(q^2)$
with the help of
\begin{eqnarray}
  \Pi^\delta(q^2) &=& \bar{\Pi}^\delta(q^2) - \frac{3}{16\pi^2}\left(\bar{C}^\delta_0 +
  \frac{\bar{C}^\delta_{-1}}{z}\right)
  \,.
\end{eqnarray}

\begin{figure}[t]
  \centering
  \includegraphics[width=.7\linewidth]{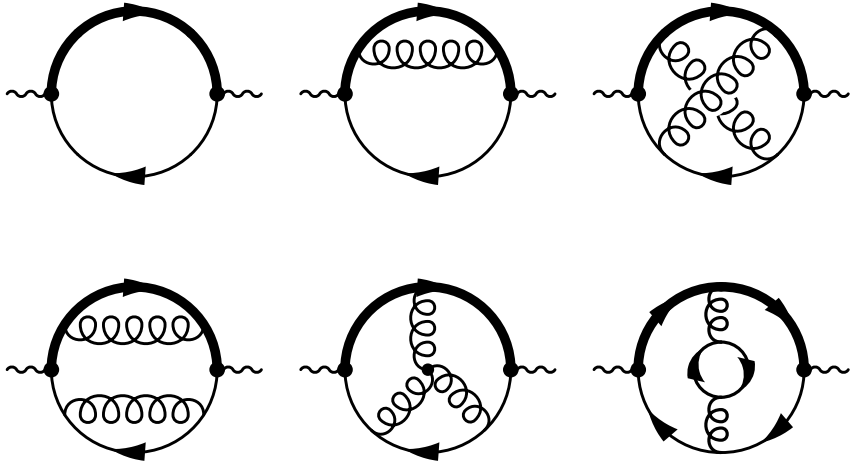}
  \caption[]{\label{fig::diags}Sample diagrams contributing to $\Pi^\delta$ at
    one, two and three loops.  The thick (upper) and thin (lower) lines
    correspond to quarks with mass $m_1$ and $m_2$, respectively, and the
    curly lines represent gluons.}
\end{figure}

Sample Feynman diagrams occurring at one-, two- and three-loop order
are shown in Fig.~\ref{fig::diags}.

\section{Master integrals for non-diagonal correlators}
\label{sec:calculation}

For the evaluation of the Feynman diagrams a well-tested chain of programs has
been used which works hand-in-hand in order to avoid error-prone manual
interactions. Since the reduction to master integrals is nowadays pretty
standard we restrict ourselves to a brief description up to this step. The
Feynman diagrams are generated with {\tt QGRAF}~\cite{Nogueira:1991ex} and the
various diagram topologies are identified and transformed to {\tt
  FORM}~\cite{Vermaseren:2000nd,Kuipers:2012rf} notation with the help of {\tt q2e} and {\tt
  exp}~\cite{Harlander:1997zb,Seidensticker:1999bb}.  In a next step
appropriate projectors are applied,
the expansion in the external momentum is performed and traces are
taken. Afterwards the scalar integrals are mapped
to functions which are then passed to {\tt FIRE}~\cite{Smirnov:2008iw} where
the reduction to master integrals is performed.

\begin{figure}[t]
  \centering
    \includegraphics[bb=130 310 430 550]{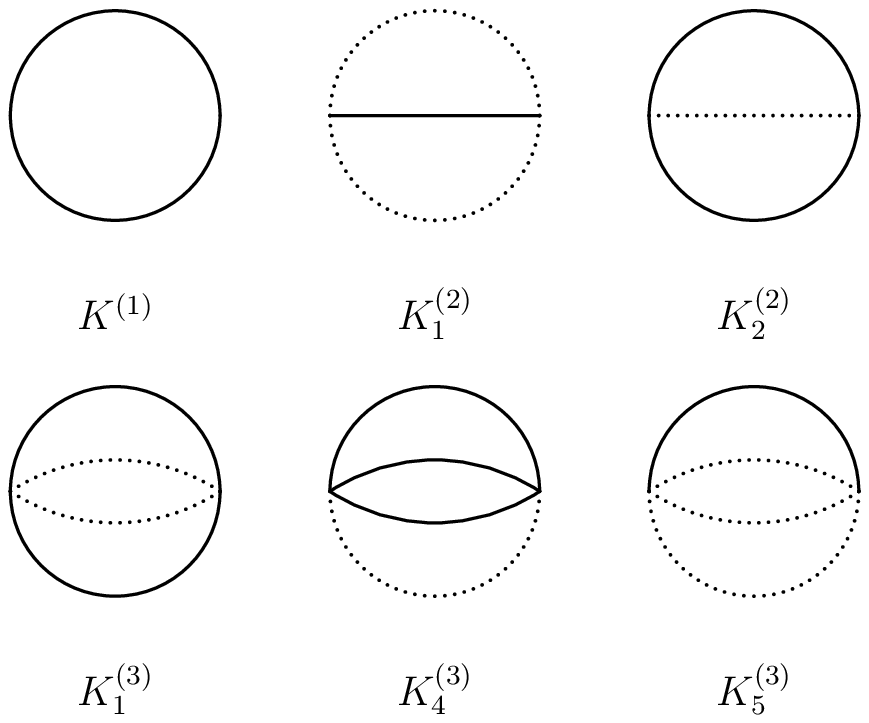}
    \\[-3em]
  \caption{One-scale master integrals where
    solid and dotted lines denote massive and massless
    propagators, respectively.}
  \label{fig:masters-one}
\end{figure}

\begin{figure}[t]
  \centering
    \includegraphics[bb=80 160 480 720]{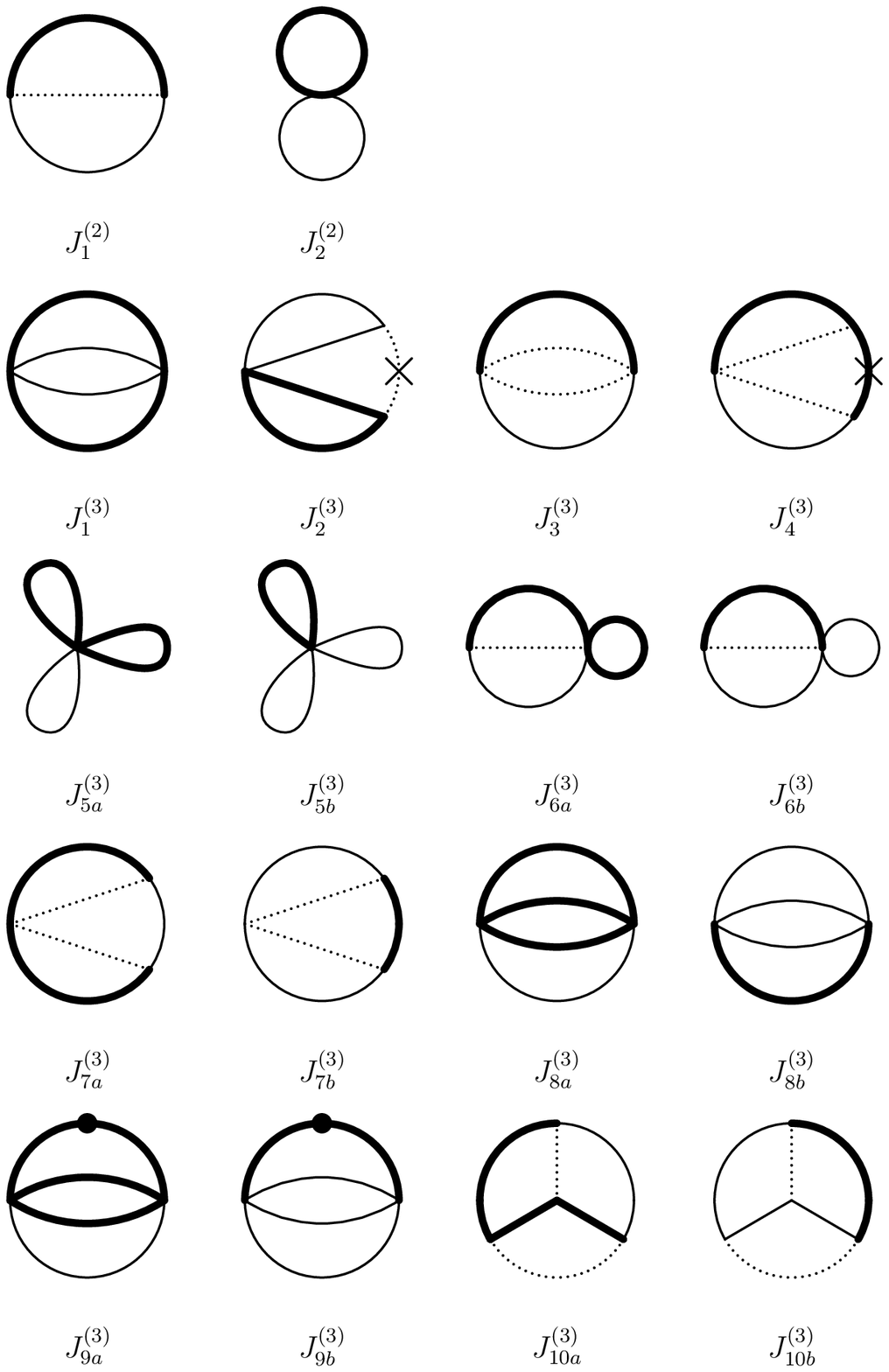}
  \caption{Two-scale master integrals where
    thick, thin and dotted lines denote heavy, light and
    massless particles. A dot on a thick line indicates that the corresponding
  propagator is squared and a cross marks propagators which are raised to
  power minus one.}
  \label{fig:masters-two}
\end{figure}

The master integrals which are needed for our calculation are shown in
Figs.~\ref{fig:masters-one} and~\ref{fig:masters-two}.  At one- and two-loop
order there are one and three master integrals, respectively, while at
three-loop order the reduction leads to three single-scale
(c.f. Fig.~\ref{fig:masters-one}) and 16 two-scale master
(c.f. Fig.~\ref{fig:masters-two}) integrals. All one- and two-loop integrals
and the single-scale three-loop integrals in Fig.~\ref{fig:masters-one} are
available in the literature (see, e.g.,
Ref.~\cite{Davydychev:1992mt,Schroder:2005va}).  For the three-loop integrals
in Fig.~\ref{fig:masters-two} only partial results exist. In this work these
results are checked by independent calculations and results for all integrals
are presented. Since the problem at hand is symmetric under the exchange of
$m_1$ and $m_2$ most master integrals come in two variants which are marked by
the subscripts ``a'' and ``b''.

\subsection{Differential equation method}

Most of the master integrals can be calculated with
the method of differential equations (see, e.g.,
Refs.~\cite{Kotikov:1991pm,Remiddi:1997ny,Caffo:1998du,Gehrmann:1999as,Argeri:2007up,Bekavac:2009gz}). In the following we
exemplify the method for
the two-loop master integral $J^{(2)}_1$ of Fig.~\ref{fig:masters-two}
which is given by
\begin{align}
  J_1^{(2)}(x) &= \left(\frac{\mu^2}{m_1^2} \right)^{4-d} m_1^2 \int
  \d^d k_1 \d^d k_2 \frac{1}{(1-k_1^2)(x^2-k_2^2)\left[-(k_1-k_2)^2\right]}
  \,,
  \label{eq::I2}
\end{align}
where $d=4-2\epsilon$ is the space-time dimension.
The differential equations are derived by
taking the derivative with respect to $x$ and reducing the resulting
integrals to master integrals. This leads to
\begin{align}
  \frac{\d}{\d x} J_1^{(2)}(x) &= -2x \frac{d-3}{1-x^2} J_1^{(2)}(x) -
  \frac{d-2}{x(1-x^2)} K^{(1)}(1) K^{(1)}(x)
  \,,
  \label{eq:1}
\end{align}
where $K^{(1)}(x)$ denotes the one-loop integral in Fig.~\ref{fig:masters-one}
with mass $x m_1$ which is given by
\begin{equation}
  \label{eq:5}
  K^{(1)}(x) = {\cal N} \frac{4}{(d-4)(d-2)} \left(x m_1\right)^2 \,,
\end{equation}
with the normalization
\begin{align}
  \label{eq:7}
  \mathcal{N} &= \left(\frac{\mu^2}{m_1^2}\right)^\ep i \pi^{d/2} \Gamma(3-d/2)
  \,.
\end{align}
To solve the differential equation it is necessary to know the initial
condition at either $x=0$ or $x=1$, which corresponds to the single-scale
integrals $K^{(2)}_1$ and $K^{(2)}_2$ shown in
Fig.~\ref{fig:masters-one}. In this simple example it is
possible to solve the differential equation without expanding in
$\epsilon$ with the result
\begin{equation}
\begin{split}
  \label{eq:6}
  J^{(2)}_1(x) =& {\cal N}^2\,m_1^2\,
  \Bigg[\frac{2 (1-x^2)^{1-2 \epsilon}
    \Gamma(-\epsilon) \Gamma(-2+2 \epsilon)}{\Gamma(1+\epsilon)} \\ &
  -\frac{x^{2-2\epsilon}(1-x^2)^{1-2\epsilon}}{(-1+\epsilon)^2
    \epsilon^2}  \ _2F_1
  \left(2-2\epsilon,1-\epsilon,2-\epsilon,x^2\right) \Bigg]\,.
\end{split}
\end{equation}
This result agrees with the one given in Ref.~\cite{Davydychev:1992mt}.
In general it is not possible to obtain a closed solution as in
Eq.~(\ref{eq:6}). In those cases one has to expand the original
differential equation in $\epsilon$ and thus obtains simpler differential
equations for the corresponding coefficients. In our example one gets from
Eq.~(\ref{eq:1}) for the first three coefficients
\begin{align}
  \frac{\dd}{\dd x} J^{(2)}_{1,-2}(x) &= \frac{2x}{x^2 - 1}
  J^{(2)}_{1,-2}(x) + \frac{2x}{x^2 - 1} \,, \nonumber\\
  \frac{\dd}{\dd x} J^{(2)}_{1,-1}(x) &= \frac{2x}{x^2 - 1}
  J^{(2)}_{1,-1}(x) - \frac{4 x}{x^2 - 1} J^{(2)}_{1,-2}(x) + \frac{2 x (1
    - 2 \log x)}{x^2 - 1} \,, \nonumber\\
  \frac{\dd}{\dd x} J^{(2)}_{1,0}(x) &= \frac{2x}{x^2 - 1}
  J^{(2)}_{1,0}(x) - \frac{4 x}{x^2 - 1} J^{(2)}_{1,-1}(x) + \frac{2 x (1
    - 2 \log x + 2 \log^2 x )}{x^2 - 1} \,,
  \label{eq::dge_coefs}
\end{align}
where the $J^{(2)}_{1,n}(x)$ denotes the coefficient of the $n^{\rm th}$ order
in the expansion in $\epsilon$.  One observes that the structure of the
homogeneous part remains the same whereas the inhomogeneous contribution
becomes successively more complicated.  While solving the differential
equations it is advantageous to use Harmonic Polylogarithms
(HPLs)~\cite{Remiddi:1999ew}, which form a special class of functions suitable
for these kind of problems. It is particularly convenient to use the
implementation in {\tt Mathematica} from
Ref.~\cite{Maitre:2005uu,Maitre:2007kp} which allows the application of the
method of the variation of constants in order to solve the differential
equation in Eq.~(\ref{eq::dge_coefs}). For our
example\footnote{Eq.~(\ref{eq::dge_coefs}) can be expressed in terms of HPLs
  with the help of $\H{0}{x}=\log x$ and $\H{0,0}{x}=(\log x)^2/2$.} at hand the
result looks as follows
\begin{equation}
  \begin{split}
    J_{1}^{(2)}(x) &= \mathcal{N}^2 m_1^2 \Bigg[ \frac{1}{\ep^2} \left(
      -\frac{1}{2} \left( x^2 + 1 \right) \right) + \frac{1}{\ep} \left(
      -\frac{3}{2} \left(1 + x^2 \right) + 2 x^2 \H{0}{x} \right) \\[5pt]
    & \quad + \frac{\pi^2 \left( x^2 - 1 \right) - 21 \left( 1 + x^2
      \right)}{6} + 6 x^2 \H{0}{x} - 4 x^2 \H{0,0}{x}\\
    & \quad - 2 \left(1 - x^2 \right) \left( \H{1,0}{x} - \H{-1,0}{x}
    \right) + \mathcal{O}(\ep) \Bigg]\,,
  \end{split}
  \label{eq::J21}
\end{equation}
which can also be obtained by expanding Eq. (\ref{eq:6}) and agrees
with Ref.~\cite{Davydychev:1992mt}.

\subsection{$J^{(3)}_{5a}$, $J^{(3)}_{5b}$, $J^{(3)}_{6a}$ and $J^{(3)}_{6b}$}

Some of the three-loop master integrals with two masses ($J^{(3)}_{5a}$,
$J^{(3)}_{5b}$, $J^{(3)}_{6a}$, $J^{(3)}_{6b}$) are products of one- and two-loop
integrals and can therefore easily be calculated with the result
\begin{equation}
  \label{eq:9}
  \begin{split}
    &J^{(3)}_{5a}(x) = \left[ K^{(1)} (1) \right] ^2 K^{(1)} (x)\,, \\
    &J^{(3)}_{5b}(x) = K^{(1)} (1) \left[ K^{(1)} (x) \right]^2\,,  \\
    &J^{(3)}_{6a}(x) = J^{(2)}_1 (x)  K^{(1)} (1)\,,  \\
    &J^{(3)}_{6b}(x) = J^{(2)}_1 (x) K^{(1)} (x)\,,
  \end{split}
\end{equation}
where $K^{(1)}(x)$ and $J^{(2)}_1(x)$ are given in Eqs.~(\ref{eq:5})
and~(\ref{eq::J21}), respectively.

\subsection{$J^{(3)}_{1}$, $J^{(3)}_{2}$, $J^{(3)}_{3}$, $J^{(3)}_{4}$,
  $J^{(3)}_{7a}$ and $J^{(3)}_{7b}$}

The method of differential equations described above can immediately be
applied to the the three-loop integrals $J^{(3)}_{1}$, $J^{(3)}_{2}$,
$J^{(3)}_{3}$, $J^{(3)}_{4}$, $J^{(3)}_{7a}$ and
$J^{(3)}_{7b}$. $J^{(3)}_{1}$ and $J^{(3)}_{2}$ obey a system of coupled
differential equations, as do $J^{(3)}_{3}$ and $J^{(3)}_{4}$. The
integrals $J^{(3)}_{7a}$ and $J^{(3)}_{7b}$ which are symmetric under
interchange of $m_1$ and $m_2$ can be obtained by solving a single
differential equation.  Solving the differential equations leads to analytical
results in terms of HPLs. The integrals needed as initial conditions are shown
in Fig.~\ref{fig:masters-one}.  The result for $J^{(3)}_{1}$ is given in
Ref.~\cite{Bekavac:2009gz}. In this reference also the result for an integral
related to $J^{(3)}_{2}$ by integration-by-parts identities is listed.  We
find agreement including terms of order $\epsilon$.

To our knowledge the results for $J^{(3)}_{3}$, $J^{(3)}_{4}$, $J^{(3)}_{7a}$
and $J^{(3)}_{7b}$ are new. They read
\begin{align}
  J_{3}^{(3)}(x) = & \mathcal{N}^3 m_1^4 \Bigg[ \frac{x^2}{3 \ep^3} +
  \frac{1}{\ep^2} \left( \frac{1}{12} \left(-1 + 16 x^2 - x^4 \right) -
    x^2 \H{0}{x} \right) \nonumber\\
  & + \frac{1}{\ep} \left( -\frac{5}{24} \left( 3 - 16 x^2 + 3 x^4
    \right) + \frac{1}{2} x^2 \left( -8 + x^2 \right) \H{0}{x} + 2 x^2
    \H{0,0}{x} \right) \nonumber\\
  & + \frac{1}{48} \left( -145 \left( 1 + x^4 \right) + 4 \pi^2 \left(
      -1 + x^4 \right) + 8 x^2 \left( 35 + 4 \zeta_3 \right) \right)
  \nonumber\\
  & + \frac{1}{12} x^2 \left( -4 \left( 30 + \pi^2 \right) + 45 x^2
  \right) \H{0}{x} - x^2 \left( -8 + x^2 \right) \H{0,0}{x} - 4 x^2
  \H{0,0,0}{x} \nonumber\\
  & -\left( 1 - x^4 \right) \left( \H{1,0}{x} - \H{-1,0}{x} \right) - 4
  x^2 \left( \H{2,0}{x} - \H{-2,0}{x} \right) + \mathcal{O}(\ep) \Bigg]
  \,,
  \\
  J_{4}^{(3)}(x) &= \mathcal{N}^3 m_1^6 \Bigg[ \frac{x^2}{3 \ep^3} +
  \frac{1}{\ep^2} \left( \frac{1}{36} \left( -2 + 45 x^2 - 6 x^4 + x^6
    \right) - x^2 \H{0}{x} \right) \nonumber\\
  &\quad + \frac{1}{\ep} \Bigg( \frac{1}{24} \left( -10 + 69 x^2 - 26
    x^4 + 5 x^6 \right) \nonumber\\
  &\qquad + x^2 \left( -4 + x^2 - \frac{x^4}{6} \right) \H{0}{x} + 2 x^2
  \H{0,0}{x} \Bigg) \nonumber\\
  &\quad + \frac{1}{72} \left( -145 - 4 \pi^2 \right) + \frac{1}{48}
  \left( 203 - 4 \pi^2 + 32 \zeta_3 \right) x^2 + \left( -\frac{37}{8} +
    \frac{\pi^2}{6} \right) x^4 \nonumber\\
  &\quad + \frac{1}{144} \left( 145 - 4 \pi^2 \right) x^6 -\frac{1}{12}
  x^2 \left( 124 + 4 \pi^2 - 82 x^2 + 15 x^4 \right) \H{0}{x} \nonumber\\
  &\quad + \frac{1}{3} x^2 \left( 24 - 6 x^2 + x^4 \right) \H{0,0}{x} -
  4 x^2 \H{0,0,0}{x} - 4 x^2 \left( \H{2,0}{x} - \H{-2,0}{x} \right)
  \nonumber\\
  &\quad - \frac{1}{3} \left( 2 + 3 x^2 - 6 x^4 + x^6 \right) \left(
    \H{1,0}{x} - \H{-1,0}{x} \right) + \mathcal{O}(\ep) \Bigg] \,,
  \\
  J_{7a}^{(3)}(x) = & \mathcal{N}^3 m_1^2 \Bigg[ \frac{1}{\ep^3} \left(
    \frac{-1 - x^2}{3} \right) + \frac{1}{\ep^2} \left( -2 - \frac{5 x^2}{3}
    + 2 x^2 \H{0}{x} \right) \nonumber\\[10pt]
  & + \frac{1}{\ep} \left( \frac{1}{3} \left( -25 - 17 x^2 + \pi^2
      \left( -1 + x^2 \right) \right) + 10 x^2 \H{0}{x} \right. \nonumber\\
  & \quad \quad \left. - 4 x^2 \H{0,0}{x} - 4 \left( 1 - x^2 \right)
    \left( \H{1,0}{x} - \H{-1,0}{x} \right) \vphantom{\frac{}{}} \right)
  \nonumber\\[10pt]
  & + \frac{1}{3} \left( -90 + 5 \pi^2 \left( -1 + x^2 \right) + 22
    \zeta_3 - 7 x^2 \left( 7 + 2 \zeta_3 \right) \right) + 34 x^2 \H{0}{x}
  \nonumber \\
  & - 20 x^2 \H{0,0}{x} + 8 x^2 \H{0,0,0}{x} + \frac{\pi^2 \left( 1 - 4
      x^2 + 3 x^4 \right)}{3 x^2} \left( \H{1}{x} - \H{-1}{x} \right)
  \nonumber\\
  & - 20 \left( 1 - x^2 \right) \left( \H{1,0}{x} - \H{-1,0}{x} \right)
  + 8 \left( 1 - x^2 \right) \left( \H{1,0,0}{x} - \H{-1,0,0}{x} \right)
  \nonumber\\
  & + 4 \left( -4 + \frac{1}{x^2} + 3 x^2 \right) \left( \H{1,1,0}{x} +
    \H{-1,-1,0}{x} \right) \nonumber\\
  & + \left( 16 - \frac{4}{x^2} - 12 x^2 \right) \left( \H{1,-1,0}{x} +
    \H{-1,1,0}{x} \right) + \mathcal{O}(\ep) \Bigg] \,.
\end{align}
$J^{(3)}_{7b}$ can be obtained from $J^{(3)}_{7a}$ by interchanging $m_1$ and
$m_2$. We have performed explicit calculations for $J^{(3)}_{7a}$ and
$J^{(3)}_{7b}$ and have used the symmetry relation as a check. Note that the
HPLs exhibit cuts along the positive real axis and thus one has to be careful
to use the proper analytic continuation.  The analytic results for all master
integrals can be found in the file {\tt TwoMassTadpoles.m} which can be
obtained from Ref.~\cite{progdata}.

There are more checks to verify the obtained results. First of all we have
checked that the results for the master integrals satisfy the original
differential equations.  Furthermore, since it is sufficient to use the value of the
integral at $x=0$ or $x=1$ as initial condition for the differential equation
the value at the other boundary can be used as cross check. A further check
constitutes the successful comparison to the expansions around $x=0$
and $x=1$~\cite{Hoff:2011ge}.

\subsection{$J^{(3)}_{8a}$, $J^{(3)}_{8b}$, $J^{(3)}_{9a}$ and $J^{(3)}_{9b}$}

The integrals $J^{(3)}_{8a}$ and $J^{(3)}_{9a}$ (and correspondingly
$J^{(3)}_{8b}$ and $J^{(3)}_{9b}$) obey a system of two coupled differential
equations which could not be solved analytically using the method described
above. Providing initial conditions a numerical solution is possible, however,
the achieved accuracy for the master integrals is not sufficient to compute
higher order moments of the current correlators since there are large
numerical cancellations between contributions from different master integrals.
Thus we decided to apply the Mellin-Barnes
method~\cite{Smirnov:1999gc,Tausk:1999vh} which provides for a given value of
$x$ a high-precision numerical result with about 30 significant digits.
Considering the integral $J^{(3)}_{8a}$ and allowing for generic indices $n_1$
and $n_2$ on a line with mass $m_1$ and $m_2$, respectively, one can derive
the two-dimensional Mellin-Barnes representation
\begin{eqnarray}
  J(n_1,n_2) &=& \int_{-i\infty}^{i \infty} \dd z_1
  \int_{-i\infty}^{i\infty}  \dd z_2 \, 
  m_1^{-2 \left(n_1+n_2+z_2+3 \epsilon -4\right)} 
  m_2^{2 z_2} 
  \pi ^{4-3 \epsilon }  \Gamma
  \left(-z_1\right) \Gamma \left(-z_2\right)
  \nonumber\\ &&\mbox{}
  \Gamma \left(-z_1-\epsilon +1\right)
  \Gamma \left(-n_1-z_1-2 \epsilon +3\right)
  \Gamma \left(-n_1-z_1-\epsilon +2\right)
  \nonumber\\ &&\mbox{}
  \Gamma \left(-n_2-z_2-\epsilon +2\right)
  \Gamma \left(n_1+n_2+z_1+z_2+2 \epsilon   -3\right)
  \nonumber\\ &&\mbox{}
  \Gamma \left(n_1+n_2+z_1+z_2+3 \epsilon -4\right) /
  \big[4\Gamma (2-\epsilon )
    \Gamma \left(n_1\right)
    \Gamma \left(n_2\right)
    \nonumber\\ &&\mbox{}
    \Gamma \left(-n_1-2 z_1-2 \epsilon +3\right)
    \big]
  \,,
  \label{eq:2}
\end{eqnarray}
where the integration contour is chosen to separate the poles from the
Gamma functions of the form $\Gamma(z_1+z_2+\ldots)$ from the ones
originating from $\Gamma(-z_1+\ldots)$ or $\Gamma(-z_2+\ldots)$.  We
used the package {\tt MB}~\cite{Czakon:2005rk} to evaluate the
Mellin-Barnes representation numerically for $0.1 \leq x \leq 1$ in
steps of $0.005$ with high precision.\footnote{For $x<0.1$ we can safely
  rely on expansions, see also Section~\ref{eq::subJ310}.  Actually, for
  the practical evaluation of the master integrals $J^{(3)}_{8a}$,
  $J^{(3)}_{8b}$, $J^{(3)}_{9a}$ and $J^{(3)}_{9b}$ we use the expansion
  around 0 for $x<0.2$ and the one around 1 for $x>0.5$.} The results
can be interpolated to obtain results for all values of $x$.  An
important cross check of the numerical approach constitutes the
comparison of the exact analytical results for the poles which can be
obtained from asymptotic expansions or the Mellin-Barnes
representation. They are given by
\begin{align}
  J_{8a}^{(3)}  &=  \mathcal{N}^3 m_1^4 \Bigg[ \frac{1}{\ep^3}\left(1+x^2\right)+\frac{1}{\ep^2}\left(\frac{15}{4}+4
    x^2-\frac{x^4}{12}-3 x^2 \log(x)\right) \nonumber\\
    &\quad + \frac{1}{\ep} \left(\frac{65}{8} +10
    x^2-\frac{5}{8} x^4+(-12 x^2 +\frac{1}{2} x^4 )\log(x)+3 x^2
    \log^2(x) \right)
  \nonumber\\&\quad + \mathcal{O}(\ep^0) \Bigg]\,, \nonumber\\
  J_{8b}^{(3)}  &= \mathcal{N}^3 m_1^4 \Bigg[ \frac{1}{\ep^3} \left(x^2+x^4\right)+\frac{1}{\ep^2} \left(-\frac{1}{12}+4
    x^2+\frac{15 x^4}{4}-3 x^2 \log(x)
    \right.\nonumber\\&\left.\quad -6 x^4 \log(x)\right)
    + \frac{1}{\ep} \left(\frac{-5}{8}+10 x^2+\frac{65}{8} x^4 -(12 x^2
    +\frac{45}{2} x^4 )\log(x)
  \right.\nonumber\\&\left.\quad +(3 x^2 +18 x^4 ) \log^2(x)\right)
+ \mathcal{O}(\ep^0)  \Bigg]\,, \nonumber\\
  J_{9a}^{(3)}  &= \mathcal{N}^3 m_1^2 \Bigg[ \frac{1}{\ep^3}
    \left(-\frac{2}{3}-\frac{x^2}{3}\right)+\frac{1}{\ep^2} \left( -\frac{3}{2}-\frac{5
      x^2}{6}+x^2 \log(x) \right) \nonumber\\
    & \quad +\frac{1}{\ep} \left(-\frac{5}{3}-\frac{4
      x^2}{3}+2 x^2 \log(x)-x^2 \log^2(x) \right)+ \mathcal{O}(\ep^0)
    \Bigg]\,, \nonumber\\
  J_{9b}^{(3)}  &=\mathcal{N}^3 m_1^2 \Bigg[\frac{-x^2}{\ep^3}+\frac{1}{\ep^2} \left(\frac{1}{6}-\frac{5
      x^2}{2}+3 x^2 \log(x)\right) \nonumber\\
    & \quad +\frac{1}{\ep} \left(1-4 x^2+6 x^2
    \log(x)-3 x^2 \log^2(x)\right)+ \mathcal{O}(\ep^0) \Bigg]\,.
\end{align}
Note that for these integrals the order $\epsilon^1$ terms are not needed for
the final results.

\subsection{\label{eq::subJ310}$J^{(3)}_{10a}$ and $J^{(3)}_{10b}$}

The differential equations for $J^{(3)}_{10a}$ and $J^{(3)}_{10b}$ only
contain inhomogeneous parts involving the integrals $J^{(3)}_{8a}$,
$J^{(3)}_{9a}$ and $J^{(3)}_{8b}$, $J^{(3)}_{9b}$, respectively, and can
therefore not be solved analytically. However, in this case the
straightforward numerical solution of the differential equations would
lead to results which are sufficiently precise for the moments
considered in this paper.

\begin{figure}[t]
  \centering
  \begin{tabular}{cc}
    \includegraphics[width=0.5\linewidth]{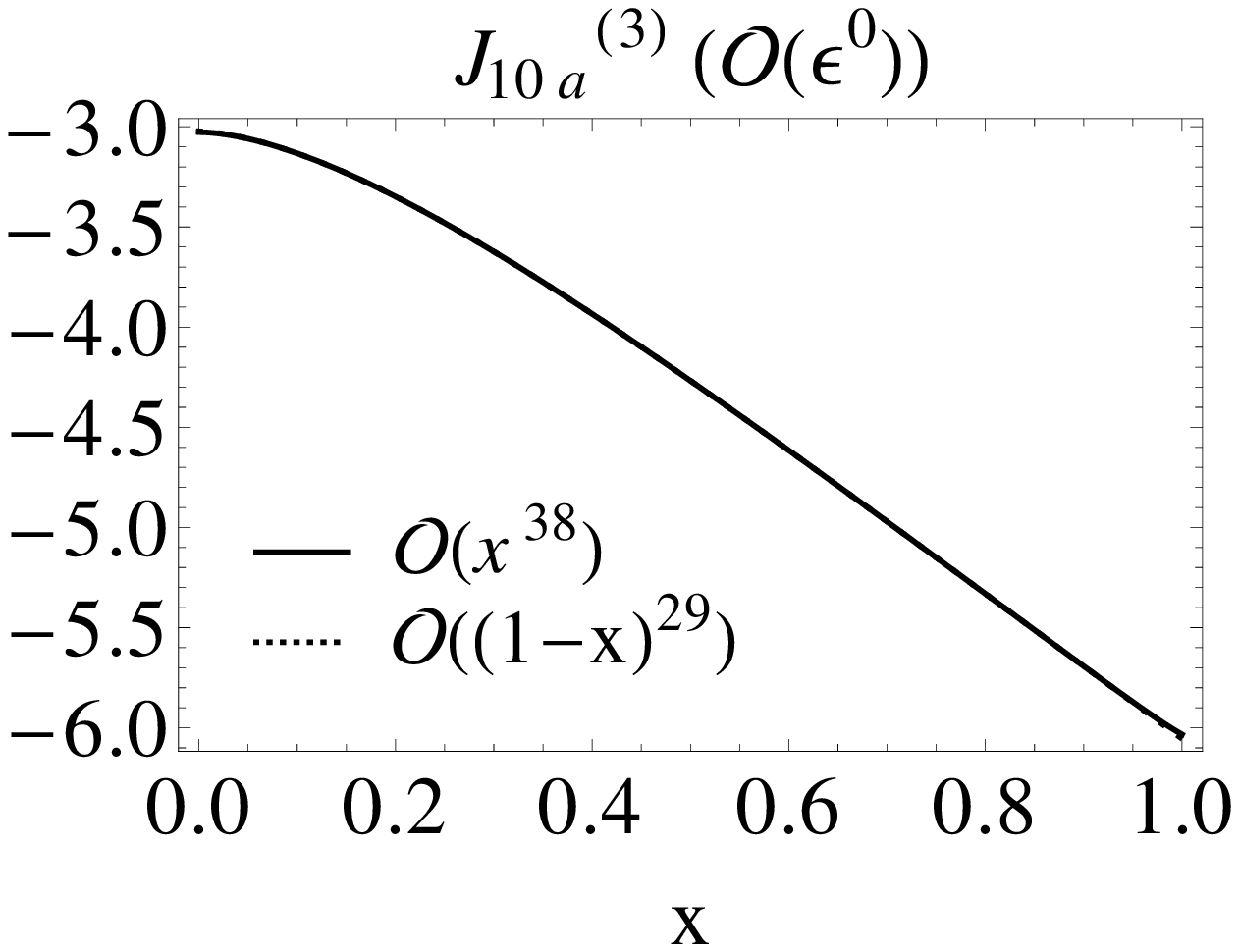} &
    \includegraphics[width=0.5\linewidth]{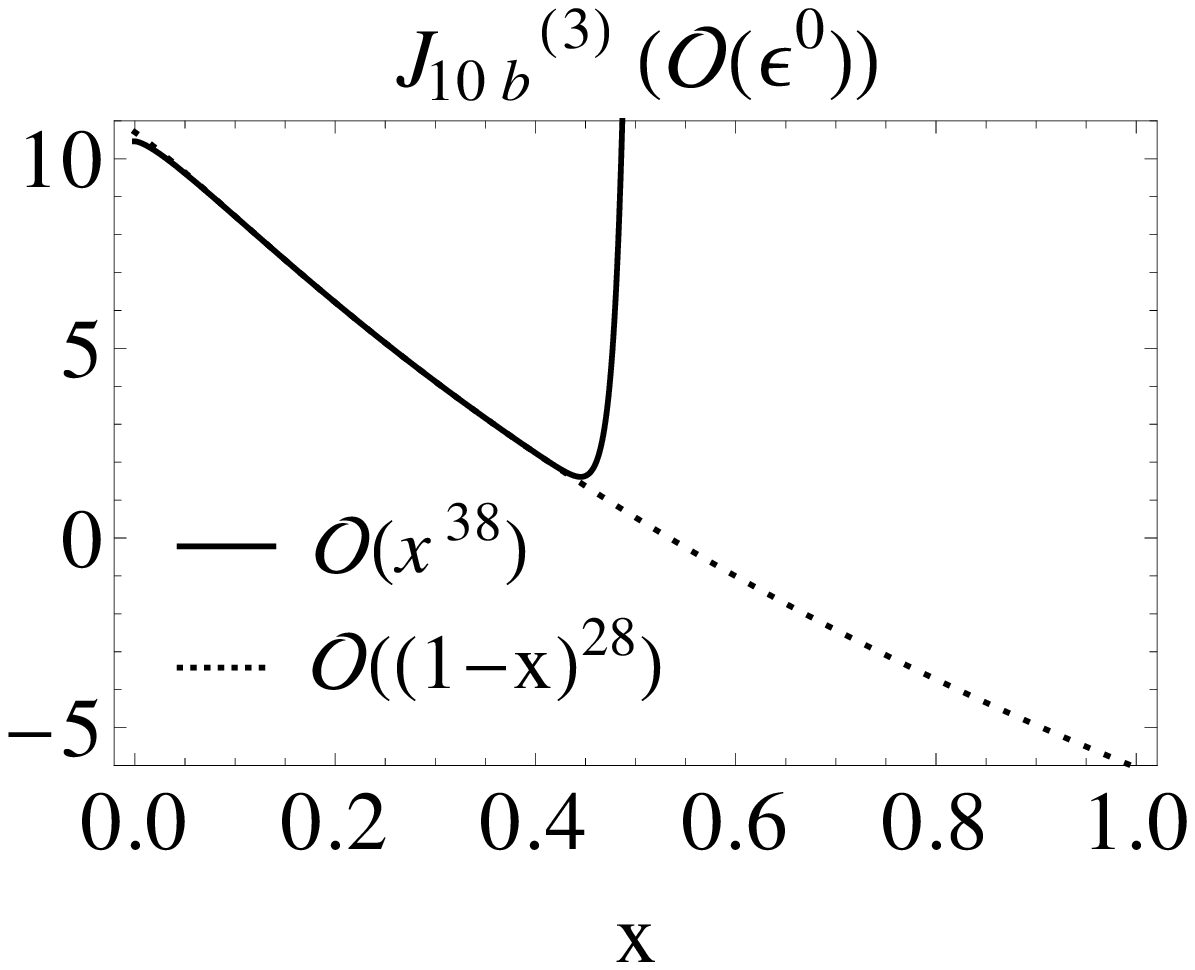} \\
        \mbox{}\hspace*{-2em}\mbox{}
    \includegraphics[width=0.45\linewidth]{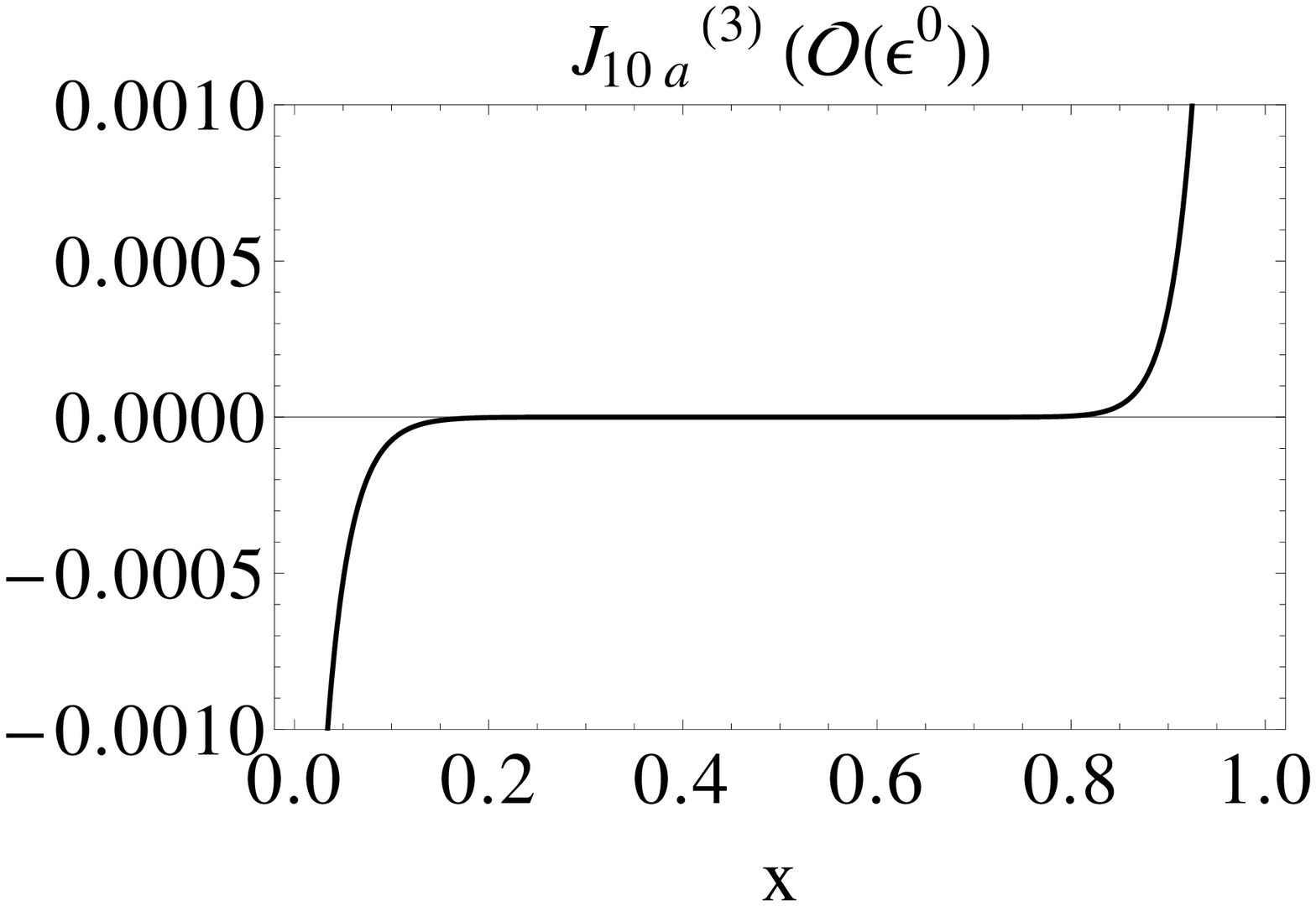} &
        \mbox{}\hspace*{-4em}\mbox{}
    \includegraphics[width=0.45\linewidth]{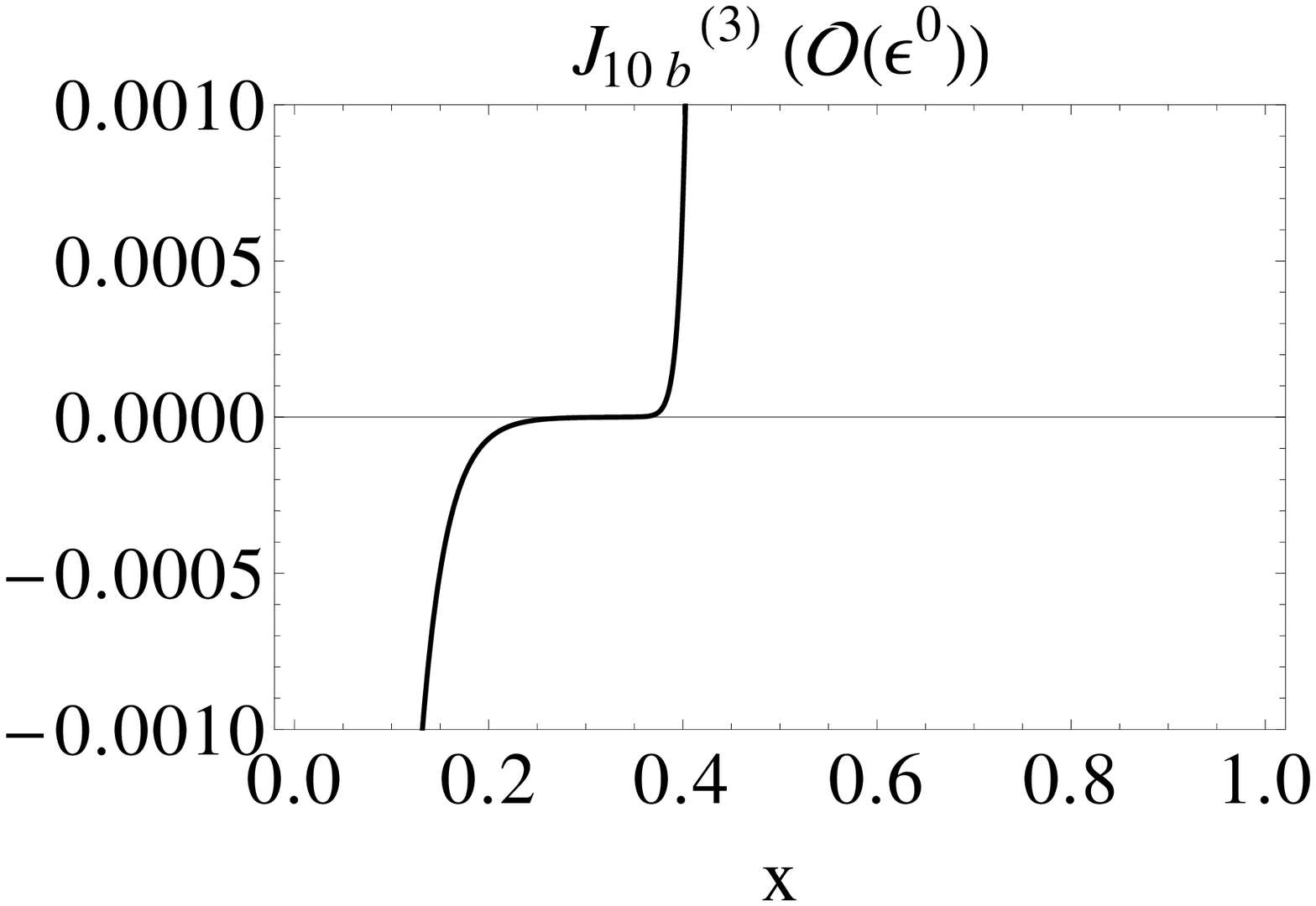}
  \end{tabular}
  \caption{Comparison of asymptotic expansions for $J^{(3)}_{10a}$ (left) and
    $J^{(3)}_{10b}$ (right).  The plots on the top show the ${\cal
      O}(\epsilon^0)$ results for the master integral where the dotted and
    solid curve corresponds to the expansion around $x=1$ and $x=0$,
    respectively. In the case of $J^{(3)}_{10a}$ the dotted curve is barely
    visible since it is on top of the solid one.  The plots on the bottom show
    the difference of the two expansions.  }
  \label{fig:J10}
\end{figure}

A more precise result for $J^{(3)}_{10a}$ and $J^{(3)}_{10b}$ which is
furthermore simpler to handle can be obtained with the help of expansions
around $x=0$ and $x=1$. In contrast to Ref.~\cite{Hoff:2011ge} where
expansions of the whole correlator have been considered and thus the expansion
depth was limited to 8 and 9 terms, respectively, in this paper
results up to order $x^{38}$ and $(1-x)^{28}$ could be obtained for the
scalar integrals $J^{(3)}_{10a}$ and $J^{(3)}_{10b}$.\footnote{Note
  that there are cancellations when constructing the physical
  moments using the expansion around $x=1$. Thus some of the moments
  are only known up to order $(1-x)^{15}$.}
The upper plots in Fig.~\ref{fig:J10}
show the results for $J^{(3)}_{10a}$ and $J^{(3)}_{10b}$ using the
expansions around $x=0$ (solid) and $x=1$ (dotted).
The difference of these results is shown in the
bottom plots.
One observes perfect agreement over a wide range of $x$ for
$J^{(3)}_{10a}$. In the case of $J^{(3)}_{10b}$ good agreement is only
found for $0.2 \le x \le 0.4$ which suggests to use the
$x\to 0$ results for $x<0.3$ and the $x\to 1$ expressions for $x>0.3$.
This has been implemented in the program for the numerical evaluation
of the master integrals.

The divergent parts of $J^{(3)}_{10a}$ and $J^{(3)}_{10b}$
can be obtained in analytical form with the help of an
asymptotic expansion. They read
\begin{eqnarray}
  J^{(3)}_{10a} &=& {\cal N}^3\,\frac{2\zeta_3}{\epsilon} + {\cal O}(\epsilon^0)\,, \\
  J^{(3)}_{10b} &=& {\cal N}^3\,\frac{2\zeta_3}{\epsilon} + {\cal O}(\epsilon^0)\,.
  \label{eq:10}
\end{eqnarray}
Also for these integrals the order $\epsilon^1$ terms do not enter the final
results. Thus numerical expressions are only needed for the finite
contributions to the moments whereas the cancellation of the poles can be
checked analytically.

All available information about the master integrals is provided in the above
mentioned {\tt Mathematica} file {\tt TwoMassTadpoles.m}~\cite{progdata}.


\section{Results}
\label{sec:results}


\begin{figure}
  \centering
    \includegraphics[bb=100 320 500 580]{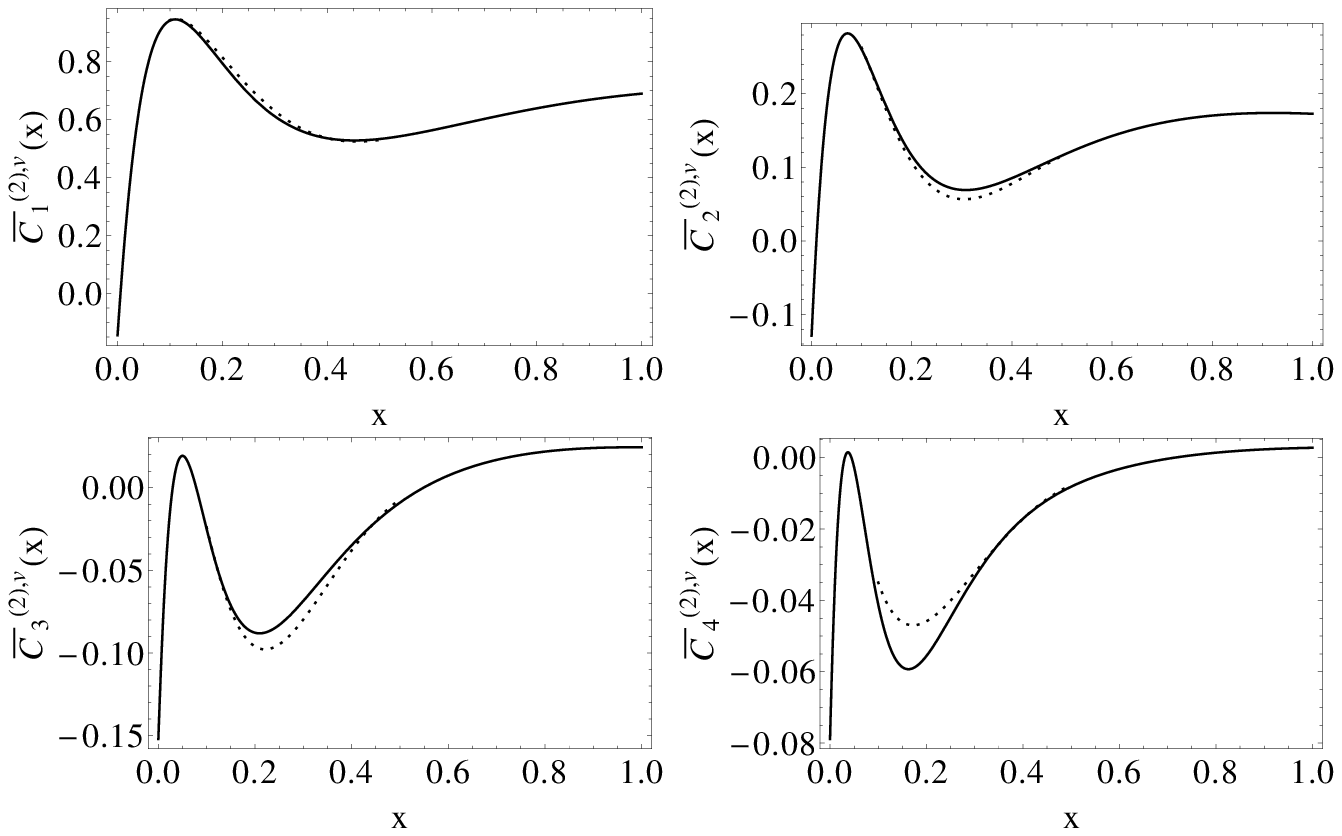}
  \caption{\label{fig:vector} Dependence on $x$ of the first four moments of
    the vector current correlator.  The dotted lines correspond to the
    interpolation results obtained in Ref.~\cite{Hoff:2011ge}.}
\end{figure}

\begin{figure}
  \centering
    \includegraphics[bb=100 320 500 580]{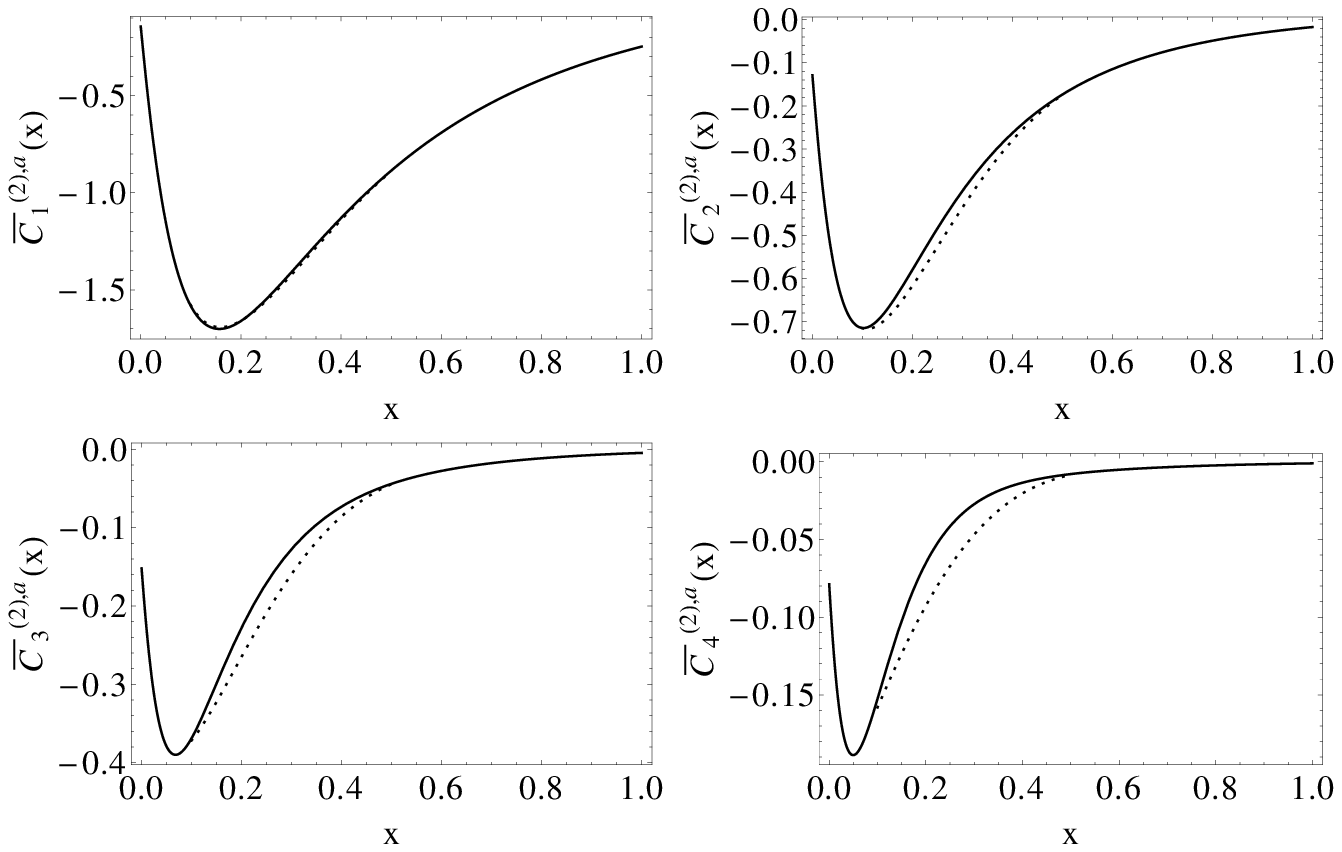}
  \caption{\label{fig:axial-vector} Dependence on $x$ of the first four moments of
    the axial-vector current correlator.  The dotted lines correspond to the
    interpolation results obtained in Ref.~\cite{Hoff:2011ge}.}
\end{figure}

\begin{figure}
  \centering
    \includegraphics[bb=100 320 500 580]{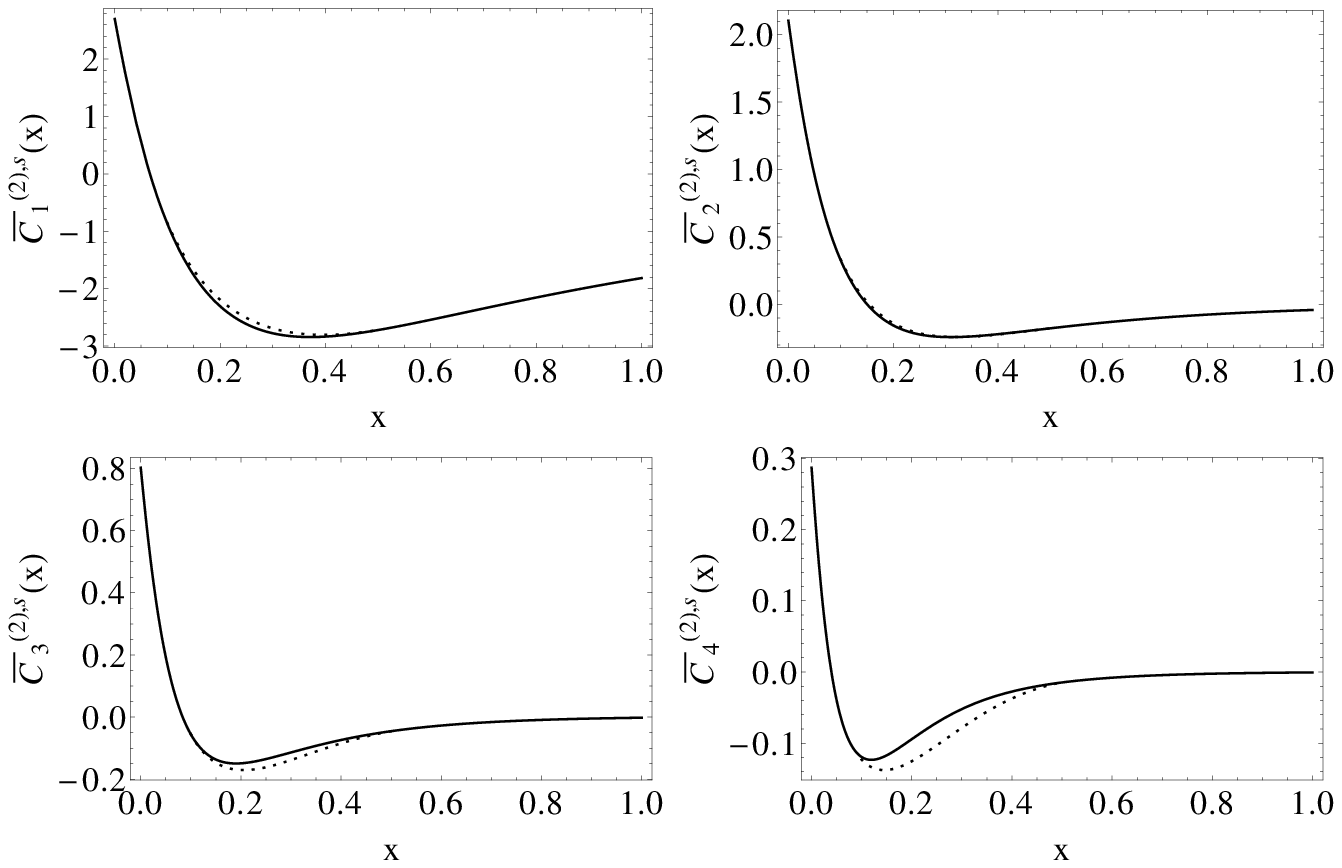}
  \caption{\label{fig:scalar} Dependence on $x$ of the first four moments of
    the scalar current correlator.  The dotted lines correspond to the
    interpolation results obtained in Ref.~\cite{Hoff:2011ge}.}
\end{figure}

\begin{figure}
  \centering
    \includegraphics[bb=100 320 500 580]{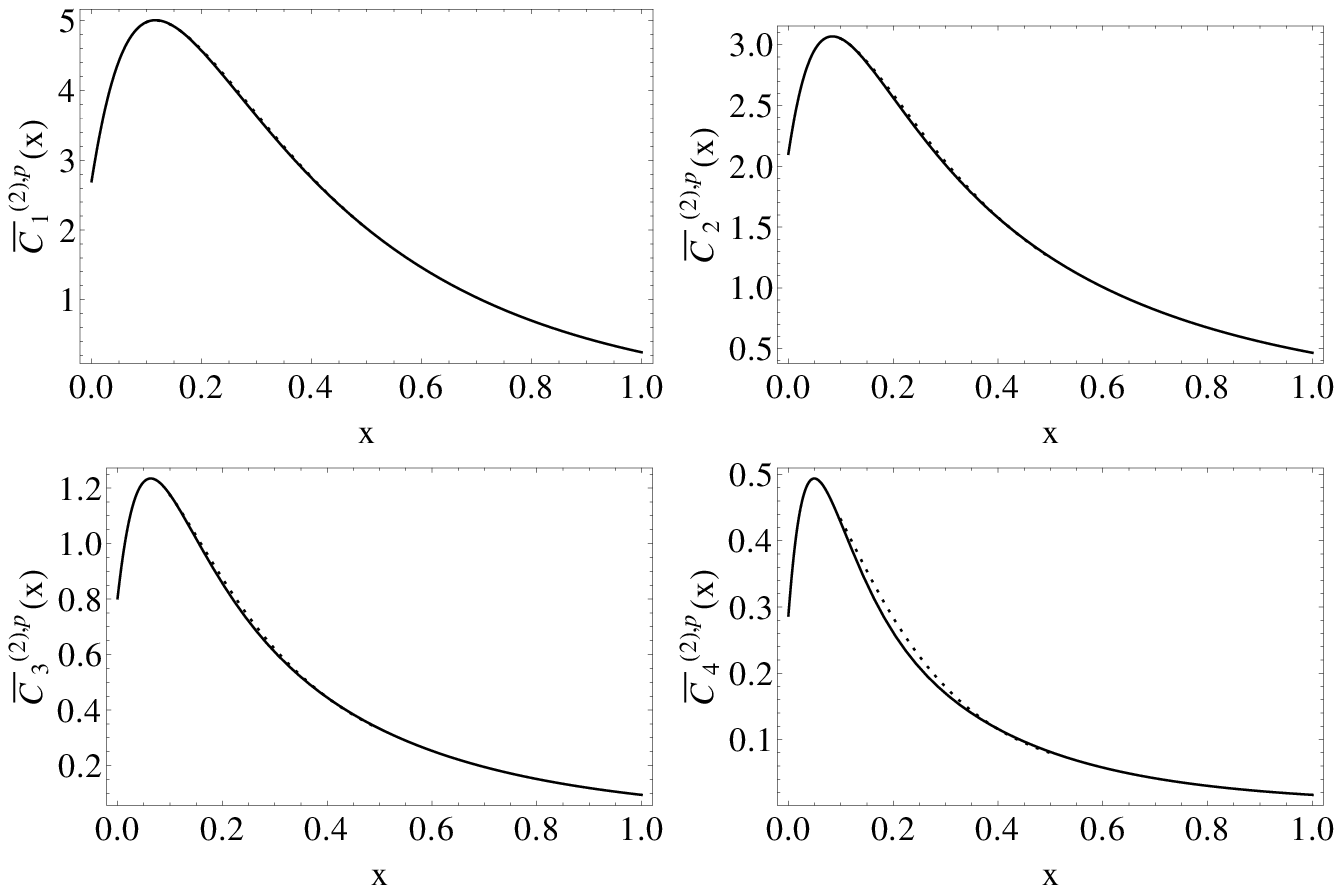}
  \caption{\label{fig:pseudo-scalar} Dependence on $x$ of the first four moments of
    the pseudo-scalar current correlator.  The dotted lines correspond to the
    interpolation results obtained in Ref.~\cite{Hoff:2011ge}.}
\end{figure}

\subsection{Moments}

In this Section we present numerical results for the first four moments of the
vector, axial-vector, scalar and pseudo-scalar current correlators.  We adapt
the colour factors to $SU(3)$ and set the number of massless quarks to
three. Furthermore, we set the renormalization scale to $m_1$.  More general
results that still contain the invariants of $SU(N_c)$ and labels for the
light and heavy quarks can be found in the {\tt Mathematica} file {\tt
  coefhl2.m}~\cite{progdata}.

For comparison we also show in Figs.~\ref{fig:vector}
to~\ref{fig:pseudo-scalar} the results obtained in Ref.~\cite{Hoff:2011ge}
as dashed lines. In Ref.~\cite{Hoff:2011ge} expansions around
$x=0$ and $x=1$ have been computed which showed good convergence properties up
to $x\lsim 0.1$ and $x\gsim 0.5$, respectively. Inbetween an interpolation has
been performed.

In most cases good agreement is found. Small deviations for $x\approx0.2
\ldots 0.3$ are observed for $\bar{C}_4^{(2),v}$ and, to a lesser extent, also
for $\bar{C}_3^{(2),v}$, $\bar{C}_3^{(2),a}$, $\bar{C}_4^{(2),a}$, and
$\bar{C}_4^{(2),s}$.  They demonstrate the importance of the calculation
performed in this paper.

From the webpage~\cite{progdata} it is possible to download a {\tt Mathematica}
file which provides the results for all moments of Figs.~\ref{fig:vector}
to~\ref{fig:pseudo-scalar}. Furthermore, also the results for
$\bar{C}_n^{(2),\delta}$ with $n=-1$ and $n=0$
and for $\bar{C}_{L,n}^v$ and $\bar{C}_{L,n}^a$ with $n=-1,\ldots,4$
are included. The one- and two-loop expressions are available up to $n=9$.


\subsection{$\rho$ parameter}
\label{sec:rho-parameter}

In this section we discuss an immediate application of $C^\delta_{-1}$ of the vector
and axial-vector current, namely the QCD corrections to the electroweak $\rho$
para\-meter which is given by
\begin{eqnarray}
  \rho = 1 + \delta\rho\,,
\end{eqnarray}
with
\begin{eqnarray}
  \delta\rho = \frac{\Sigma_Z(0)}{M_Z^2} - \frac{\Sigma_W(0)}{M_W^2}
  \,.
\end{eqnarray}
$\Sigma_W(0)$ and $\Sigma_Z(0)$ are the transverse parts of the $W$ and $Z$
boson propagators defined through
\begin{eqnarray}
  \Sigma_{W/Z}(0) &=& \frac{g_{\mu\nu}}{d} \, \Pi_{W/Z}^{\mu\nu}\,,
\end{eqnarray}
where $\Pi_{W/Z}^{\mu\nu}$ are the corresponding polarization functions.  The
QCD corrections to $\delta\rho$ in the Standard Model (SM), where all quark
masses except the top quark mass are set to zero, are known up to four
loops~\cite{Veltman:1977kh,Djouadi:1987di,Chetyrkin:1995ix,Avdeev:1994db,Schroder:2005db,Chetyrkin:2006bj,Boughezal:2006xk}. In
the following we consider a generic fourth generation of quarks which couples
to the $W$ and $Z$ boson in the same way as the top and bottom quarks of the
SM. For simplicity we denote the new doublet by $(t^\prime,b^\prime)$.

\begin{figure}[t]
  \centering
  \begin{tabular}{cc}
    \includegraphics[width=.4\linewidth]{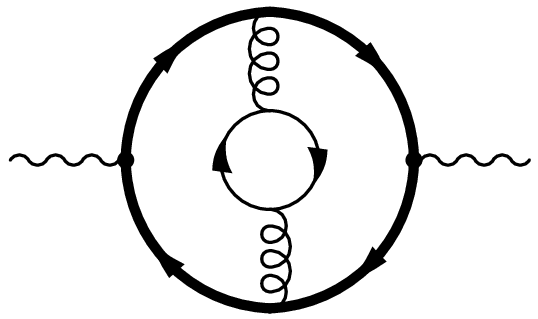}
    &
    \raisebox{.35em}{
      \includegraphics[width=.45\linewidth]{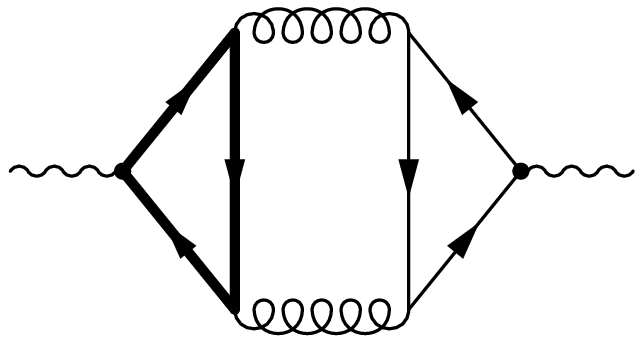}
    }
    \\
    (a) & (b)
  \end{tabular}
  \caption[]{\label{fig::db_sing}Additional contributions to the diagonal
    correlator with two mass scales: (a) double bubble and (b) singlet
    contribution.}
\end{figure}

$\Sigma_W(0)$ can be obtained from $\bar{C}_{-1}^v$ and $\bar{C}_{-1}^a$
via the relation
\begin{eqnarray}
  \frac{\Sigma_W(0)}{M_W^2} &=& \frac{3 G_F m_{t^\prime}^2}{16\pi^2\sqrt{2}}
  \left( \bar{C}_{-1}^v(x) + \bar{C}_{-1}^a(x) \right)
  \,.
\end{eqnarray}
In the case of $\Sigma_Z(0)$ there is no contribution form the vector
part. Furthermore one has to consider additional contributions at three-loop
order which are only present for currents where $\psi_1=\psi_2$ in
Eq.~(\ref{eq::currents}). Thus we cast the $Z$-boson self energy in the form
\begin{align}
  \frac{\Sigma_Z(0)}{M_Z^2} &= \frac{3 G_F m_{t^\prime}^2}{16 \pi^2 \sqrt{2}}
  \left(\bar{C}_{-1,\rm{diag}}^a + \left(\frac{\alpha_s}{\pi}\right)^2
    \delta \bar{C}_{-1,\rm{db}}^{a,(2)} +
    \left(\frac{\alpha_s}{\pi}\right)^2 \delta
    \bar{C}_{-1,\rm{sing}}^{a,(2)} \right)\,,
  \label{eq::sigZ}
\end{align}
where the quantities $\delta\bar{C}_{-1,\rm db}^{a,(2)}$ and
$\delta\bar{C}_{-1,\rm sing}^{a,(2)}$ receive contributions from the
double bubble and singlet Feynman diagrams involving two mass scales
(see Fig.~\ref{fig::db_sing}). The double bubble diagrams involving only one
massive quark are contained in $\bar{C}_{-1,\rm{diag}}^a$
together with all other one-scale contributions.
Note that the singlet-type contributions involving massless
up, down, strange and charm quarks add up to zero.

In Appendix~\ref{app::db_sing} explicit analytical results are presented
for $\bar{C}_{-1,\rm{diag}}^a$, $\delta\bar{C}_{-1,\rm db}^{a,(2)}$ and
$\delta\bar{C}_{-1,\rm sing}^{a,(2)}$.  Let us mention that the
treatment of $\gamma_5$ for the singlet diagrams can be found in
Ref.~\cite{Larin:1993tq}.

Before presenting results we transform the $\overline{\rm MS}$ quark masses
$m_{t^\prime}$ and $m_{b^\prime}$ to the on-shell
scheme~\cite{Gray:1990yh,Bekavac:2007tk} since this is common practice in the
context of analyzing electroweak precision data. For $M_{b^\prime}=0$ we
reproduce the SM result~\cite{Chetyrkin:1995ix,Avdeev:1994db}.
The general result can be cast in the form
\begin{align}
  \delta \rho_{\textrm{OS}} &= \frac{3 G_F M_{t^\prime}^2}{16
    \pi^2 \sqrt{2}} \Bigg\{
  2 (1 + X^2) + \frac{8 \, X^2}{1 - X^2} \, \H{0}{X}
  +  \frac{\alpha_s}{\pi} \Delta^{(1)}(X)
  +  \left(\frac{\alpha_s}{\pi}\right)^2 \Delta^{(2)}(X)
  \Bigg\}
  \,,
\end{align}
with $X=M_{b^\prime}/M_{t^\prime}$ being the ratio of the on-shell quark
masses.
The two-loop result reads
\begin{align}
  \Delta^{(1)} &=
    \frac{16}{3} \left\lbrack %
    - \frac{3}{12} \left( 1 + X^2 \right) %
    + \frac{X^2}{1 - X^2} \, \H{0}{X} %
    + \frac{2 X^2 \, \left( 3 + X^4 \right)}{\left( 1 - X^2 \right)^2} \,
    \H{0,0}{X} %
    \right.\nonumber\\&\left.
    - \left( 1 - X^2 \right) \left( \frac{\pi^2}{12} + H_{1, 0} (X) -
      H_{-1, 0} (X) \right) %
  \right\rbrack %
  \,,
\end{align}
which agrees with the findings of Ref.~\cite{Djouadi:1993qv}.
Using the asymptotically expanded master integrals we obtain a series
expansion in $X$ and $1-X$. The results including terms of order $X^2$ and
$(1-X)^4$ read
\begin{align}
  \Delta^{(2)}(X) &= \frac{85}{324}-\frac{D_3}{9}-\frac{2845 \pi
    ^2}{486}+\frac{26 \pi ^4}{135}+\frac{441 S_2}{4}-\frac{4}{9} \pi ^2
  l_2+\frac{4}{27} \pi ^2
  l_2^2
  \nonumber\\ &\quad
  -\frac{4 l_2^4}{27}-\frac{32}{9} a_4 -\frac{664 \zeta_3}{27} +n_l
  \left(-\frac{1}{9}+\frac{13 \pi
      ^2}{27}-\frac{8 \zeta_3}{9}\right)
  - X \ \frac{2 \pi ^2}{3} +X^2 \Bigg[\frac{1943}{81}
  \nonumber\\ &\quad
  +\frac{535 \pi ^2}{972}-\frac{\pi
    ^4}{6}+\frac{1125 S_2}{4}-\frac{4}{9} \pi ^2 l_2+\frac{4}{27} \pi ^2
  l_2^2-\frac{4 l_2^4}{27}
  +\frac{1154  l_X^2}{9} -\frac{304 l_X^3}{9}
  \nonumber\\&\quad
  -\frac{32}{9} a_4+l_X
  \left(\frac{82}{9}-\frac{32 \pi ^2}{9}+\frac{128}{9} \pi ^2 l_2 -16 \pi ^2
    l_2-\frac{4
      \zeta_3}{3}\right)
  +n_l \left(-\frac{5}{3}+\frac{7 \pi^2}{27}
    \right.\nonumber\\ &\left. \quad
  +\left(-\frac{4}{3}+\frac{8 \pi ^2}{9}\right) l_X-\frac{52
    l_X^2}{9}+\frac{32 l_X^3}{9}+\frac{8 \zeta_3}{9}\right)-\frac{1651
  \zeta_3}{27}\Bigg]
  + \mathcal{O}(X^3)
\,,
\end{align}
where $l_X=\log X$ and $\mu^2=M_{t^\prime}^2$ has been
adopted\footnote{Results for general $\mu$ can be found in {\tt
    rho.m}~\cite{progdata}.}  and
\begin{eqnarray}
  a_4&=&\mbox{Li}_4(1/2)\,\,\approx\,\, 0.51747906167389938633\,,
  \nonumber\\
  S_2&=&\frac{4}{9\sqrt{3}}\mbox{Im}(\mbox{Li}_2(e^{i\pi/3}))
  \,\,\approx\,\,0.26043413763216209896\,,\nonumber\\
  D_3&=&6\zeta_3 -\frac{15}{4}\zeta_4  -6
  [\mbox{Im}(\mbox{Li}_2(e^{i\pi/3}))]^2
  \,\,\approx\,\,-3.0270094939876520198\,.
\end{eqnarray}
Similarly, the expansion around $M_{b^\prime}\approx M_{t^\prime}$ leads to
\begin{align}
  \Delta^{(2)}(X) &= (1-X)^2 \left[
    -\frac{5933}{108}
    -\frac{16}{27} \pi ^2
    l_2+\frac{2807 \zeta_3}{72}
    +n_l \left(-\frac{38}{27}+\frac{8 \pi ^2}{27}\right)
  \right]
  \nonumber\\
  &\quad + (1-X)^3 \left(-\frac{116}{9}+\frac{8 n_l}{9}\right)
  \nonumber\\
  &\quad +(1-X)^4 \left[
    -\frac{465547}{4860}+\frac{4 \pi ^2}{135}
    +\frac{4}{135} \pi ^2
    l_2+\frac{157781 \zeta_3}{2160}
    +n_l\left(\frac{473}{810}-\frac{2 \pi ^2}{135}\right)
  \right] \nonumber\\
&\quad + \mathcal{O}((1-X)^5)\,.
  \label{eq::rhoexp2}
\end{align}
In the file {\tt rho.m}~\cite{progdata} analytical results up to order
$X^{30}$ and $(1-X)^{20}$ are provided.

In the above equations $n_l$ labels the number of massless quark flavours.  In
order to reproduce the SM result with massless bottom quark one has $n_l=4$
and $x=0$.  The contribution of a fourth generation is obtained for $n_l=6$
which assumes a massless top quark at three-loop order.

In the SM the higher order corrections to the $\rho$ parameter are quite
important. E.g., the three-loop corrections correspond to a shift in the top
quark mass of about 2~GeV which is larger than the current experimental
uncertainty. In the limit of equal quark masses $\delta\rho$ is proportional
to the mass difference and thus the numerical impact of the higher order
corrections is reduced as can be seen from Eq.~(\ref{eq::rhoexp2}).



\section{Summary}
\label{sec::sum}

In this paper moments of correlators formed by vector, axial-vector, scalar
and pseudo-scalar currents are considered to three-loop order. The currents
couple to quarks with different masses $m_1$ and $m_2$ which leads to
two-scale vacuum integrals after expanding in the external momentum of the
correlators. The moments can be expressed as a linear combination of 12
non-trivial three-loop master integrals which are discussed in details in
Section~\ref{sec:calculation}. In particular, a {\tt Mathematica} package {\tt
  TwoMassTadpoles.m} is provided~\cite{progdata} which contains analytical
results for all but six master integrals. For the latter high-precision
numerical results are available.  In a further {\tt Mathematica} file {\tt
  coefhl2.m} we also provide results for the first four moments for each
correlator.

As a by-product we compute three-loop corrections to the $\rho$
parameter for a generic fourth generation of quarks with masses
$m_{t^\prime}$ and $m_{b^\prime}$ for the up- and down-type quark. In
addition to the calculation for the moments of the non-diagonal
correlators one has to compute double-bubble and singlet-type diagrams
where two different masses can be present in the fermion triangles. For
the latter analytical results are provided; the complete results for the
$\rho$ parameter up to three loops is available in {\tt
  rho.m}~\cite{progdata}.



\section*{Acknowledgments}

This work was supported by the BMBF through Grant No. 05H09VKE
and the Graduierten\-kolleg ``Elementarteilchenphysik bei
h\"ochster Energie und h\"ochster Pr\"azision''.


\begin{appendix}


\section{\label{app::db_sing}Appendix: Explicit results for
  $\bar{C}_{-1,\rm{diag}}^a$,
  $\delta\bar{C}_{-1,\rm db}^{a,(2)}$ and $\delta\bar{C}_{-1,\rm
    sing}^{a,(2)}$}

In this appendix analytical results are provided for the building blocks of
$\Sigma_Z(0)$ in Eq.~(\ref{eq::sigZ}) using the $\overline{\rm MS}$ definition
of the quark masses.  For convenience we choose for the renormalization
scale $\mu^2=m_1^2$; general results are available in electronic form from
Ref.~\cite{progdata}.
Defining
\begin{eqnarray}
  \bar{C}_{-1,\rm{diag}}^a &=& \bar{C}_{-1,\rm{diag}}^{a,(0)}
  +\frac{\alpha_s}{\pi}\bar{C}_{-1,\rm{diag}}^{a,(1)}
  +\left(\frac{\alpha_s}{\pi}\right)^2\bar{C}_{-1,\rm{diag}}^{a,(2)}
  \,,
\end{eqnarray}
we have
\begin{align}
  \bar{C}_{-1,\rm{diag}}^{a,(0)} &= -\frac{4 \left(1+x^2\right)}{\ep}+8
  x^2 \H{0}{x} \,, \nonumber\\
  \bar{C}_{-1,\rm{diag}}^{a,(1)} &= \frac{4 \left(1+x^2\right)}{\ep^2}
  -\frac{10 \left(1+x^2\right)}{3 \ep}+ \frac{1}{3} \left(-1-x^2\right)
  -\frac{8}{3} x^2 \H{0}{x}-32 x^2 \H{0,0}{x} \,, \nonumber\\
  \bar{C}_{-1,\rm{diag}}^{a,(2)} &= \frac{1}{\ep^3} \left(-\frac{19}{3}
    \left(1+x^2\right)\right)+\frac{1}{\ep^2} \left(\frac{281}{18}
    \left(1+x^2\right)+\frac{8}{3} x^2 \H{0}{x}\right) \nonumber\\
  &\quad +\frac{1}{\ep} \Big(-\frac{4}{3} x^2 \H{0}{x}-16 x^2
  \H{0,0}{x}-\frac{1}{54} \left(1+x^2\right) \left(401+6 \pi ^2
  \right.\nonumber\\
  &\left.\quad -36 \zeta_3\right)\Big) +\frac{1}{18} x^2 \H{0}{x}
  \left(-181+12 \pi ^2-72 \zeta_3\right) -\frac{272}{9} x^2 \H{0,0}{x}
  \nonumber\\
  &\quad +\frac{1136}{3} x^2 \H{0,0,0}{x} -\frac{1}{90}
  \left(1+x^2\right) \Big( -790 + \left( -5 + 40 l_2^2 \right) \pi^2 +
  \frac{22}{3} \pi^4 - 40 l_2^4 \nonumber\\
  &\quad + 1160 \zeta_3 - 960 a_4 \Big) +n_l \Bigg(\frac{2
    \left(1+x^2\right)}{9 \ep^3}-\frac{5 \left(1+x^2\right)}{9
    \ep^2}+\frac{4 \left(1+x^2\right)}{9 \ep}-\frac{5}{9} x^2
  \H{0}{x}\nonumber\\
  &\quad +\frac{32}{9} x^2 \H{0,0}{x}-\frac{32}{3} x^2
  \H{0,0,0}{x}+\frac{1}{18} \left(1+x^2\right) (-1+32 \zeta_3) \Bigg) \,.
\label{eq::Cdiag}
\end{align}
The double bubble and singlet result is given by
\begin{align}
  \delta \bar{C}_{-1, \mathrm{db}}^{a,(2)} &= \frac{4
    \left(1+x^2\right)}{9 \ep^3}+\frac{-\frac{2}{9}
    \left(1+x^2\right)-\frac{8}{3} x^2 \H{0}{x}}{\ep^2} \nonumber\\
  & \quad +\frac{\frac{1}{9} \left(-1+\pi ^2\right)
    \left(1+x^2\right)+\frac{4}{3} x^2 \H{0}{x}+16 x^2 \H{0,0}{x}}{\ep}
  \nonumber\\
  & \quad +\frac{2}{3} \left(2-\left(1+\pi ^2\right) x^2\right) \H{0}{x}
  -\frac{4}{3} \left(1+7 x^2\right) \H{0,0}{x} \nonumber\\
  &\quad +\frac{2 \left(1-8 x-6 x^2-8 x^3+x^4\right) \H{-1,0,0}{x}}{3
    x}\nonumber\\
  &\quad -\frac{256}{3} x^2 \H{0,0,0}{x}+\frac{2 \left(1+8 x-6 x^2+8
      x^3+x^4\right) \H{1,0,0}{x}}{3 x}\nonumber\\
  &\quad -\frac{1}{18} \left(1+x^2\right) \left(49+\pi ^2-38 \zeta_3
  \right) \,, \nonumber\\
  \delta \bar{C}_{-1, \mathrm{sing}}^{a,(2)} &= 8 x \left( \H{1,0,0}{x} +
    \H{-1,0,0}{x} \right) - 7 \left( 1 + x^2 \right) \zeta_3 \,.
\label{eq::Cdbsing}
\end{align}
The three-loop result $\bar{C}_{-1,\rm{diag}}^{a,(2)}$ depends on the
number of massless flavours $n_l$.  Let us finally mention that the {\tt
  Mathematica} package {\tt rho.m} contains replacement rules which
express the {\tt HPL}s occurring in Eqs.~(\ref{eq::Cdiag})
and~(\ref{eq::Cdbsing}) in terms of logarithms and dilogarithms.


\end{appendix}



\end{document}